\documentclass[iop]{emulateapj}

\def \eg{{e.g.,}}
\def \etal{{et~al.\null}}

\slugcomment{Astrophysical Journal, in press}

\shorttitle{The Halo Field Stars in M81}
\shortauthors{Durrell et al.}

\begin{document}

\title{Deep HST/ACS Photometry of the M81 Halo}

\author{Patrick R. Durrell}
\affil{Department of Physics \& Astronomy, Youngstown State University,
Youngstown, OH 44555; prdurrell@ysu.edu}
    
\author{Ata Sarajedini}
\affil{Department of Astronomy, University of Florida,
    Gainesville, FL 32611}

\author{Rupali Chandar}
\affil{Department of Physics \& Astronomy, University of Toledo,
Toledo, OH 43606}
    
\begin{abstract}

We present a deep color-magnitude diagram for individual stars in the halo
of the nearby spiral galaxy M81, at a projected distance of 19~kpc, based on
data taken with the {\em Advanced Camera for Surveys} on the {\em
Hubble Space Telescope} ($HST$).  The CMD reveals a red giant branch
that is narrow and fairly blue, and a horizontal branch that has stars
that lie mostly redward of the RR Lyrae instability strip.  We derive
a mean metallicity of [M/H] = --1.15 $\pm$ 0.11 and age of 9 $\pm$ 2
Gyr for the dominant population in our field, from the shape of the
red giant branch, the magnitude of the red clump, and the location of
the red giant branch bump.  We compare our metallicity and age results
with those found previously for stars in different locations within
M81, and in the spheroids of other nearby galaxies.

\end{abstract}


\section{Introduction}

Spheroidal stellar populations, such as elliptical galaxies and the
halos and bulges of spiral galaxies, contain a significant fraction of
all stars in the local universe.  The ages and metallicities of halo
stars in these galaxies provide some of the most important
observational clues to the formation and earliest evolution of
galaxies, whether through studies of globular star clusters (GCs) or, in
more nearby galaxies, the individual halo stars.  Below, we broadly
summarize what is currently known about the ages and metallicities of
stars in the spheroids of nearby galaxies.

The spheroid of the Milky Way has two components, a centrally
concentrated bulge composed of more metal-rich stars, and an extended,
lower density halo consisting of more metal-poor stars. The halo of
the Galaxy shows substructure, with (at least) two chemically distinct
components.  Stars of the `inner halo' have a mean [Fe/H]$\sim -1.6$
\citep[\eg][]{rn91,ap06,iv08}, while stars in the `outer halo' (R$>$15 kpc)
are more metal-poor, with a mean [Fe/H]$\sim -2.2$ \citep{car07}.  The
halo also contains many stellar streams and tidal debris that are
probably the remnants of previously accreted satellite galaxies
\citep[\eg][and references therein]{newb02,yanny03,juric08,smith09,newb09,klem09}.  The globular
clusters in the Galaxy have a weak metallicity gradient inside $\ga$10
kpc, with no such trend apparent beyond 10~kpc out to the most distant
cluster located at $\approx$120 kpc
\citep{zinn85,har01}. The stars and GCs in the halo of the Milky Way appear
to be mostly old, with ages between 11 and 13~Gyr \citep{car07,
mf09}, indicating that the dominant era of accretion in the Milky Way
took place soon after formation \citep{hamm07}.

Detailed studies of the stars in M31 suggest that this galaxy has had
a more active, recent accretion history when compared with the Milky
Way.  M31 has a markedly metal-rich inner spheroid component
\citep[\eg][]{mk86,dur01,bro03,kal06,gil07} that is dominated by 
substructure \citep{fer02,ibata07,rich08}.  M31 accreted a relatively
massive satellite galaxy only $\sim1$~Gyr ago
\citep{font06,fard06,fard07,fard08} , and has had other many more
minor interactions \citep{mcc09b} as well.  Deep images taken with the
{\em HST}, which reach below the main sequence turnoff region for
ancient stars in M31, indicate the presence of metal-rich,
intermediate age stars ($\approx6-10$~Gyr) in fields located at
deprojected distances of 11, 20, and 35 kpc, with a larger
contribution from older, more metal-poor stars increasing at larger
distances \citep{bro03,bro06,bro07,bro08}.

Beyond $\sim30$~kpc and extending out to at least $\sim160$~kpc, the
halo of M31 is dominated by metal-poor stars with [Fe/H]$\sim -1.5$
\citep{irwin05,raja06,kal06,chap06}, and this component too has copious
substructure \citep{ibata07,mcc09b}.  Little is known about the ages
of these stars except for the recent work of \citet{mack09} who
studied the M31 GC MGC1, which is located at a projected
galactocentric distance of $\sim$120 kpc. The CMD of this cluster
reveals a metal abundance of [M/H]$\sim$--2.3 and a horizontal branch
morphology that is consistent with those of Milky Way GCs at this
metal abundance, suggesting a similar age.

Studies of stars in the spheroids of several more distant elliptical
galaxies suggest that metal-rich stars dominate out quite far,
probably to at least 10-15$R_{eff}$. NGC~5128 has a largely
metal-rich population at radii of 8-33~kpc \citep{har99,hh00,hh02}.  
Even further out, an analysis of the colors 
of RGB stars, which are sensitive to metallicity, and the colors and
luminosities of the AGB bump and the RC, which are sensitive to both
metallicity and age, suggests that stars in NGC~5128 at $R\sim38$~kpc
have largely intermediate to old ages, with $8\pm3$~Gyr \citep{rej05}.  
The halos of NGC~3377 and NGC~3379 have a wide range of
metallicity, with a significant fraction of metal-rich stars, although
metal-poor stars begin to dominate the spheroid of NGC~3379 beyond
$\approx12R_{eff}$.  The lack of stars brighter than the RGB suggests
that NGC~3377 and NGC~3379 contain few young, bright AGB stars \citep{har07a,har07b}.  
Like NGC~3379, a dominant population of metal-poor
stars have been found in the spheroid of the edge-on spiral galaxy
NGC~891 \citep{rej09}.

M81 is an earlier-type spiral (Sab) than either the Milky Way or M31.
However, studies of the resolved stars in the outskirts of M81 have
already shown that it too is a complex mix of stellar populations.
\citet{tik05} used 9 archival WFPC2 images of M81 to show
that the number density of RGB and AGB stars decreases with
galactocentric radius, eventually flattening out at a (deprojected
distance) of $\sim25$~kpc.  They attribute the stars interior to this
as part of the thick disk, and stars beyond this as part of the halo.
CMDs of fields interior to 25~kpc reveal relatively young stars
\citep{wil09,angst}, and older RGB stars with a relatively large range
in metallicity ($-1<[M/H]< 0$; Mouhcine \etal \ 2005, Williams \etal \ 2009).
These stars are likely located in the disk, either thin or thick, of
M81.  More recently, \citet{bark09} found a flattending of the stellar
density profile beyond $\sim20$~kpc (deprojected) from M81, with more
metal-poor stars ([M/H]$\sim-1.1$) than in the interior fields, again
indicating that the halo dominates at these distances.  The only
constraint on the {\em ages} of stars in the halo of M81 comes from
the measurement of absorption line strengths in globular clusters,
which suggest ages that are indistinguishable from the GCs in the
Galaxy, within the significant uncertainties ($\approx3-4$~Gyr)
\citep{sch02}.

In this work, we study the metallicities and ages of stars in a field
in M81 that lies at a projected distance of 19~kpc, using observations
taken with the Advanced Camera for Surveys on $HST$.  These data are
among the deepest available for a portion of M81.  The rest of
this paper is organized as follows: Section~2 presents the data and
reduction; Section~3 presents the CMD for our field and discusses
general features; Section~4 presents a new determination of the
distance to M81 from the tip of the RGB, and averages this with
previous distance determinations to give our adopted value; Section 5
discusses the red clump, and Section~6 presents an estimate of the metallicities and ages 
of stars in our M81 field.  Section~7 presents some new
results on a small population of blue stars discovered in our field,
and Section~8 discusses the results in terms of the formation of M81,
as well as our main conclusions.

\section{Observations and Reductions}

Observations of M81 were obtained with the Wide Field Channel (WFC) of
the Advanced Camera for Surveys (ACS) on-board the $HST$, for program
GO-10604 (PI: Sarajedini).  A single ACS/WFC pointing covering
$\approx3.4\arcmin \times 3.4\arcmin$ was centered at
$\alpha=9^h 53^m 03\fs 20$, $\delta= +68\degr 52\arcmin 03.6\arcsec$ (J2000.0), at a projected
distance of 18\arcmin \ from the center of M81, or 19 kpc at the
adopted distance of 3.7~Mpc (see Section 4).

The specific pointing was chosen to lie along the southwestern
semi-minor axis, beyond the radius suggested as the edge of M81's
thick disk \citep{tik05}, where the surface brightness of M81 is still
high enough that we can determine a meaningful metallicity
distribution from RGB stars.  Figure~1a shows the location of our ACS
pointing.  It also shows the WFPC2 field studied by \citet{mou05},
which falls on the edge of the suggested thick disk in M81, at a
projected distance of 12.8\arcmin \ (14 kpc) from the center of M81.
Figure~1b shows our pointing superposed on HI contours from
\citet{yun94}; while our field is outside of M81's purported thick
disk (at a deprojected distance of 30-34 kpc), it still lies along the
line-of-sight with some of the many HI filaments that exist outside
M81's optical disk \citep{yun94}.  While some of this HI has been
tidally stripped from M81 by encounters with M82 and NGC~3077,
probably $\approx 220-280$~Myr ago \citep{yun99}, velocities of the HI
towards our field are consistent with rotation of the M81 disk (M.Yun, private communication).

Our field was observed on 2005 Sept. 11-13  for 5 orbits (2
images per orbit) in the F606W filter and 9 orbits (2 images per
orbit) in the F814W filter, using a standard 4-point dither pattern.
Each individual exposure had an exposure time of 1247s, leading to a
total exposure time of 12470s in the F606W filter, and 22446s in the
F814W filter.  We retrieved the pipeline flatfielded images from the
$HST$ archive as well as the pipeline-processed multidrizzled combined
images from subsets of the data.

Photometry was performed with the ACS module of the DOLPHOT software
package \citep{dol00}, which is specifically designed for point-source
photometry of objects in the individual FLT images, which were taken
directly from the $HST$ archive.  Object detection and photometry was
performed on all exposures simultaneously, using one of our deep
$F814W$ drizzled images (derived from a subset of 8 images) as the
reference frame.  The DOLPHOT parameters were set to the recommended
values given in the DOLPHOT User's Guide
\footnote{http://purcell.as.arizona.edu/dolphot/}, which includes
PSF-fitting using pre-derived PSFs, as well as the calculation and
application of CTE correction \citep{cte03} and aperture corrections.
The instrumental magnitudes were converted to the VEGAmag $HST$
photometric system by adopting the revised zeropoints (for data taken
before July 4, 2006) of 26.420 for F606W and 25.536 for F814W, which
differ by $\sim 0.02 - 0.03$ mag from the values originally published
by \citet{sir05}.  We refer to magnitudes in this system by their
filter name ($m_{F606W}$ and $m_{F814W}$), and in cases when we need
to convert to the Johnson-Cousins systems refer to the filters as $V$
and $I$.

We determined the photometric completeness limits and associated
errors in our point-source photometry by adding and remeasuring 60,000
artificial stars in the ACS images.  The artificial stars have a range
of magnitude (24 $< m_{F814W} <$ 31) and color ($-0.25 < m_{F606W} - m_{F814W} <
2.25$).  The Pritchet interpolation function \citep{fl95}

\begin{equation}
 f(m) = {{1}\over{2}}\left( 1 - {{\alpha(m - m_{lim})}\over{\sqrt{1 + 
\alpha^2(m - m_{lim})^2}}} \right)
\end{equation}

\noindent was used to fit the binned completeness fractions $f$ ($=$
number of stars detected / number of stars added) as a function of
magnitude.  Here, $\alpha$ is a parameter that measures the rate of
decline of $f(m)$ at $m_{lim}$, and $m_{lim}$ is defined as the 50\%
completeness limit.  To account for completeness variations in both
color and magnitude, we used only the bluest stars ($m_{F606W} -
m_{F814W} =0$) to determine the F814W completeness fraction
$f(F814W)$, and redder stars ($m_{F606W} - m_{F814W} >1$) to derive
$f(F606W)$.  The limiting magnitude \citep[$f=0.5$;][]{har90} for each
filter is $m_{F606W, lim} = 29.73$ and $m_{F814W, lim} = 28.79$.

\section{General Features in the Color-Magnitude Diagram}

We selected point sources in our field as follows.  First, we
restricted our sample to objects with a DOLPHOT category consistent
with stars (OBTYPE=1).  Of the sources that met this criterion, only
those that were measured on at least 8 of the 10 individual F606W
images, and on at least 15 of the 18 individual F814W images were
retained, yielding a list of 13,858 sources over the entire ACS field.
To remove remaining, slightly resolved objects, we applied the
following cuts: $\chi^2 < 2$ (from the PSF fitting), and a crowding
parameter CROWD $< 0.4$ mag, where the latter rejected objects whose
magnitudes are likely to be affected by neighboring sources.  Finally,
we kept only those objects with sharpness values within an
empirically-derived hyperbolic function : $|SHARP|
<0.01+0.16e^{0.46(F814W-28)}$.  All of these criteria were chosen
based on visual inspection of sources on the ACS images.  The
$m_{F606W}$--$m_{F814W}$ versus $m_{F814W}$ color-magnitude diagram
(CMD) for the 9081 remaining objects (representing our `cleaned'
sample of stellar objects in the M81 field) is shown in the left panel
of Figure~2, with the CMD of all objects {\it rejected} by the
CHI/SHARP/CROWD cuts plotted in the right hand panel.  The photometric
errors derived from the artificial star experiments, as well as the
50$\%$ photometric completeness levels, are also plotted in the
left-hand panel.

Comparison of the two plots shows that the vast majority of rejected
objects do not correspond to the sequences visible in the `clean' CMD,
lending credence to our choice of classification criteria.  We note
that some true stellar sources will be removed by our somewhat
stringent selection criteria, but since we are most interested in
those objects with the best quality photometry, these are acceptable
`losses'.

The cleaned CMD extends from above the tip of the first ascent red
giant branch (TRGB) to one magnitude below the horizontal branch (HB).
Stars on the red giant branch are older than a few Gyr.  The HB has a
morphology which is predominantly redward of the RR Lyrae instability
strip, although we cannot unambiguously detect an extended blue
horizontal branch due to the presence of a small population of blue
stars ($m_{F606W}$--$m_{F814W}\sim$0), which will be discussed in
Section 7.  In fact, the red HB resembles a red clump (RC) most
commonly seen in relatively young (down to $\approx1$~Gyr) and/or
metal-rich stellar populations.  An analysis of the RGB and RC stars
is presented in Section 5.  There is a small but noticeable population
of stars above the TRGB, likely asymptotic giant branch (AGB) stars of
intermediate age ($\sim$5 to 10 Gyr). From the Besan\c{c}on Galaxy
model of \citet{rob03}, we estimate that there should be $\approx7$
(foreground) stars from the Milky Way with $21 < I <24$ and $1.5 <
~(V-I)~$$<4$ in our field of view, so the vast majority of stars in
our M81 field in this color and magnitude range belong to M81.  We will explore
these populations in more detail in the upcoming sections.

\section{DISTANCE TO M81}

\subsection{The RGB Luminosity Function}

We determine the luminosity function of the RGB stars in our field by
defining an `RGB zone' and thereby excluding stars far from the RGB
sequence.  We use the Z=0.0001 and Z=0.004 models of \citet{gir02} to
provide a general guide to the shape of the RGB in both color and
luminosity, after first making the models redder and bluer by 0.05 mag
to accomodate photometric uncertainties; see Figure 3.  However, we
explicitly include stars brighter than the tip of the RGB, likely to
be AGB stars, by allowing a larger range of magnitudes than the model
predictions.

In Figure~4, we show the $m_{F814W}$ luminosity function of the RGB in
our field for those objects in the extraction region with
$22<m_{F814W}< 28$, with a bin width of 0.08, and uncertainties given
by $\sqrt{N}$.  We do not make a correction for foreground stars,
because we found in \S3 that these are very sparse in this direction.
The red clump is clearly present at $m_{F814W}\sim 27.8$; we will
return to this feature in Section 5.

\subsection{Tip of the Red Giant Branch}

In order to interpret our CMD further, we need to know the distance to
M81.  Here, we determine the distance from the tip of the RGB (TRGB)
in the $I$-band, which has proven to be an excellent distance
indicator for metal-poor, old stellar populations \citep[\eg][]{lfm93,
sakai96, fer00, bell01, mcc04, mou05}.  Visual inspection of our CMD
shows a transition in the number counts at $m_{F814W} \sim 24.0$,
albeit with a small number of stars.   While a small number of stars in the 
upper part of the RGB (there are $\sim 110$ in our CMD) 
can lead to a possible systematic bias in the determination of 
$I_{TRGB}$, \citet{mf95} have shown that with at least $\sim 100$
stars in the top magnitude of the RGB (as is the case for our
luminosity function), such biases are expected to be small.   

To derive a quantitative estimate of the location of the RGB tip, we
have employed the Sobel edge-detection filter \citep{sakai96} in order
to find the strongest discontinuity in the bright part of our LF in
Figure 4 (for the relatively small number of stars in the upper RGB,
this simple edge-detection method should suffice for our purposes).
Each object in the LF is modelled as a Gaussian distribution with a
dispersion equal to the expected photometric error.  We use the
dispersion in the mean of the individual DOLPHOT magnitudes as our
photometric error.  The resulting luminosity function $\Phi (m)$ is
plotted at the top of Figure 5.  To determine the location of the RGB
tip, we applied to this LF the edge-detection algorithm

\begin{equation}
E(m) = \Phi (m+ \sigma(m)) - \Phi(m - \sigma(m))
\end{equation}

\noindent where $\sigma(m)$ is the mean photometric error defined from
our artificial star experiments \citep{sakai96}.  The resultant
response function is plotted in the bottom of Figure 5, and shows a
peak at $m_{F814W} = 24.00 \pm 0.02$, where the small formal
uncertainty reflects only the width of the Sobel function peak.
Converting to the $I$-band via the synthetic transformations of
\citet{sir05} (and using the revised zeropoints, as noted above), we
find $I_{TRGB} = 23.97 \pm 0.04$, where we have adopted an uncertainty
of 0.03 mag in the transformation from $m_{F814W}$ to $I$.  The true
uncertainty in $I_{TRGB}$ is almost certainly larger than that given
above, since they do not include any uncertainties in the photometry
itself.

The distance to M81 is determined by comparing the measured magnitude
for the tip of the RGB found above with the predicted intrinsic
magnitude, $M_{I,TRGB}=-4.05 \pm 0.02$ (for [Fe/H]$=-1.6$) from
\citet{riz07}.  This predicted value is very similar to
$M_{I,TRGB}=-4.06 \pm 0.07$ (random error) derived by \citet{fer00},
and to $M_{I,TRGB}= -4.04 \pm 0.12$ (for [Fe/H]$=-1.7$) from
\citet{bell01}.  A comparison between observed and predicted
$M_{I,TRGB}$ magnitudes gives an observed distance modulus of $(m-M)_I
= 28.02 \pm 0.05$ for M81, or an intrinsic value of $(m-M)_o = 27.86
\pm 0.06$ ($d = 3.73\pm 0.16$ Mpc), where we have assumed a foreground
extinction of $A_I=0.16\pm 0.04$ \citep[using $E(B-V)=0.08$;][]{sch98}
and adopted a 0.02 uncertainty in the foreground reddening to M81.
This is consistent with determinations from independent TRGB estimates
and other methods, as discussed below.

\subsection{Previous and Adopted Distance Determinations}

Several different techniques have been used previously to estimate the
distance to M81, leading to distance modulus estimates that differ by
approximately 0.1~mag.

The period-luminosity relationship for Cepheid variables is a robust
way to determine the distance to nearby galaxies.  \citet{fer00}
determined a value of $(m-M)_o = 27.80 \pm 0.08$ for M81, based on 30
Cepheids discovered in M81 by \citet{fre94}, using $HST$ observations.
More recently, \citet{mcc09} determined a distance modulus of
$27.78\pm0.05\pm0.14$ based on 11 Cepheids, in good agreement with the
distance determined by Ferrarese et al.  Here, we adopt the McCommas
\etal\  value as the Cepheid based distance to M81.

Estimates of the distance to M81 based on the tip of the RGB method
tend to give a range of values.  We found $(m-M)_o = 27.86 \pm
0.06$ above, and \citet{tik05} derived a mean value of $(m-M)_o =
27.92\pm 0.04$, based on fields which are interior to that used here.
However, \citet{riz07} derive a much shorter distance modulus of
$(m-M)_o = 27.68\pm 0.04$ from their detection of the TRGB in an M81
disk field, and \citet{angst} derived a value of $(m-M)_o =27.77$ 
from deep photometry of an ACS field in M81's disk.

For this work, we adopt a distance of
$\langle$$(m-M)_o$$\rangle$$=27.80 \pm 0.08$ ($d=3.63\pm 0.14$ Mpc) to
M81, which is the mean value of the five distance determinations
described above (4 TRGB values and the Cepheid value).  The quoted error is the 
dispersion of the input values (each of the individual distance determinations have similar 
uncertainties).  

\section{Location of the Red Clump}

Inspection of Figure 2 shows a well-populated horizontal branch,
indicating a population of relatively old stars.  While our data do
not allow for a clean detection of a blue horizontal branch, the
magnitude and color of the red clump (RC\footnote{Here we equate the
RC with the `red horizontal branch' (RHB), where the vast majority of
helium-burning stars lie redward of the instability strip}) allows us
to estimate the age and metallicity of the dominant stellar population
in our field, as shown in the next section.

To derive the photometric properties of the RC, we use only those
stars with colors in the range $0.80 < V - I < 1.10$; a close-up of
this part of the CMD is shown in Figure 6, along with the luminosity
function of stars in this color range, where the RC is clearly
visible.  To derive the mean magnitude of this feature, we fit the
`background' LF immediately brighter and fainter than the RC as a
linear function, and fit a single Gaussian to the LF once the linear
background has been subtracted off.  We note that the choice of bin
size and the exact background fit have negligible effect on the mean
magnitude determined for the RC clump.  The resulting fit (and
1$\sigma$ uncertainties) yields $I_{RC} = 27.75\pm 0.04$.  The
Gaussian width of $\sigma_{I} = 0.14\pm 0.04$, while consistent with
the expected photometric errors from the artificial star experiments,
does allow for some intrinsic spread in age and/or metallicity. Given
our adopted distance modulus of $\langle$$(m-M)_o$$\rangle$$=27.80 \pm
0.08$, the I-band absolute magnitude in this region of M81 is
$M_I(RC)$ = --0.21 $\pm$ 0.10.  The location of the RC in our dataset
is also consistent with the value derived by \citep{wil09} for their
M81 outer disk field\footnote{This may seem odd considering the large
differences in stellar population between the two
studies.  \citet{wil09} quote $m_{F814W,RC} = 27.792 \pm 0.002$ while we
get $m_{F814W,RC} = 27.78 \pm 0.04$. The Williams \etal \ field exhibits
a mean age of 2-3 Gyr and a metallicity of [M/H]$\sim -0.7$.  In our
field (see next section) we find a mean age of 9 Gyr and a [M/H]$ =
-1.15 \pm 0.03$.    Our M81 field is $\sim 6$ Gyr older and $\sim 0.5$ dex more
metal-poor than the Williams \etal field. A change of $-0.5$ dex makes
the red clump 0.15 mag brighter and a change of +6 Gyr makes the red
clump 0.17 mag fainter \citep[based on Fig. 8 of][]{wil09}, so they
offset each other resulting in a RC absolute magnitude that is
essentially unchanged.}.

We note the presence of a small positive feature at $I\sim 27.42\pm
0.05$ (adopted error based on the bin size used) in the LF in Figure
6.  This feature is very likely the faint signal from the RGB
bump\footnote{This feature is not the AGB bump, which would be more
luminous than the RC by $\approx1$~mag \citep[\eg][]{rej05,wil09}.  We
do not see strong evidence for an AGB bump here because of the
relatively low number of stars in our field.  \citet{wil09} do see an
AGB bump, because their field contains many more stars, but do not
observe an RGB bump, likely because their significantly larger range
of metallicities and ages washes out this feature.}, the location on the RGB where the hydrogen-burning
shell crosses the discontinuity in chemical abundances remaining from
the stars' convective envelope \citep{fer99,as99}.  The
luminosity of the RGB bump depends on both the metallicity and the age
of the stars, and will be used as a consistency check on our results
in the next section.

\section{Mean Metallicity and Age of the M81 Field}

The absolute magnitude of the red clump, as well as the shape (slope)
and location (color) of the RGB are both sensitive to the age and
metallicity of the stellar population.  However, the RGB is more
sensitive to metallicity than to age, and the magnitude of the RC is
more sensitive to age than to metallicity, so it is possible to use
both features together to constrain the age and metallicity of the
dominant population.

We compare the observations with the theoretical isochrones of
\citet{gir00}, assuming scaled-solar abundance.\footnote{The
determination of stellar metallicities for the RGB stars in our field
is performed in the HST VEGAmag $m_{F606W}-m_{F814W}$ CMD as this
makes use of the well-determined photometric zeropoints for the
ACS/WFC instrument. However, the theoretical dependence of the red
clump magnitude on metallicity and age is performed in the
ground-based system as it requires knowledge of the median I-band RC
magnitude for a population of stars, and these have only been
constructed for the ground-based filters. This is unlikely to
introduce a significant systematic effect because the ACS/WFC F814W
passband closely matches its ground-based $I$-band counterpart. We also
note that the RGBs of the \citet{gir00} isochrones are identical to
those of \citet{mar08}.}  We first correct the colors and magnitudes
using the values for reddening and distance noted above.  Next, we
assume an age and use the colors of RGB stars with $M_I$$<$--2.0 to
determine metallicity based on an interpolation within the
\citet{gir00} grid of theoretical isochrones with
--2.3$\leq$[M/H]$\leq$0.0.  Figure 7 shows these theoretical RGBs
overplotted on our M81 CMD for an age of 9 Gyr. The $M_I$$<$--2.0
magnitude range is chosen to minimize the effect of asymptotic giant
branch stars on the blue side of the RGB \citep[\eg][]{sj05}.  The
resultant mean metallicity and the I-band absolute magnitude of the RC
derived above, $M_I(RC)$ = --0.21 $\pm$ 0.10, are used in conjunction
with the \citet{gir00} models for the median magnitude of the red
clump stars to infer the age of the population. This age estimate
$\tau$ is then used to redetermine the mean metallicity from the RGB
stars and the process is repeated. This iterative procedure converges
quickly and yields a mean metallicity of [M/H] $= -1.15 \pm 0.03$
(random), $\pm$0.11 (systematic), where the latter value includes the
influence of errors in the distance modulus and reddening. The mean
age of the dominant population is then $\langle$$\tau$$\rangle$ = 9
$\pm$ 2 Gyr, where the quoted uncertainty reflects the error in the
red clump magnitude, which is the dominant source of uncertainty.  The
resultant metallicity distribution function (MDF) is shown in Fig. 8
wherein the fitted Gaussian profile gives a 1-$\sigma$ width of 0.39
dex in [M/H].

The dominant signal in the MDF is due to stars around $M_I \sim -2$ to
$\sim -3$. The artificial star tests we have performed suggest that
the mean color error at $M_I\sim -2.4$ is 0.032 mag, which corresponds
to $\sigma_{[M/H]} \sim$ 0.12 dex based on the \citet{gir00}
isochrones in the range --1.3$\leq$[M/H]$\leq$--0.7.  This suggests
a significant fraction of the $\sigma_{[M/H]}$ exhibited by the
MDF is due to an intrinsic range in metallicity among the M81 stars in
our field. In particular, subtracting the dispersion due to errors
from the measured dispersion in quadrature yields an intrinsic
abundance spread of $\sigma_{[M/H]}$= 0.37 dex.

It is also important to keep in mind that we have used scaled-solar
isochrones for the determination of the MDF shown in Fig. 8. The
degree to which field stars in the M81 halo are enhanced in the
$\alpha$-capture elements is currently not known. We can estimate the
effect of such an enhancement on our MDF using the formalism of
\citet{ferr00}:

\[
[M/H] = [Fe/H] + log(0.638f_{\alpha} + 0.362)
\]

\noindent where $f_{\alpha}$ represents the factor by which the
$\alpha$-elements are enhanced.   Thus, if $[\alpha/Fe]$=$+0.3$, then
$f_{\alpha}$=2 and our computed scaled-solar [M/H] values are larger than 
[Fe/H] by $\sim 0.2$ dex.

We conclude this section by providing a series of consistency checks
on our principle result, which is that the mean metallicity and age of
stars in our M81 field are [M/H]$= -1.15 \pm 0.11$ and $9 \pm 2$~Gyr,
respectively.  First, the location of the weak RGB bump (described in
the previous section) at $I\sim 27.42\pm 0.05$ can be used to provide
an independent estimate of the age/metallicity of the M81 field stars.
Indeed, the RGB bump lies above the HB/RC for primarily old, more
metal-poor populations \citep{fer99,as99}.  From our adopted distance
modulus, an assumed mean color $(V-I)=0.95\pm 0.05$ and $A_V=0.27$, we
find $M_{V,RGB bump} \sim 0.30 \pm 0.10$.  From the relations in
\citet{as99} (and adopting an additional 0.1 dex uncertainty in those
relations), this translates to a metallicity [Fe/H]$\sim -1.6\pm 0.2$
for an old, 12 Gyr population, or a metallicity [Fe/H]$\sim -1.3\pm
0.2$ for a 9 Gyr population.  While the uncertainties in these
results are large, and the $\alpha$ enhancement unknown, the age and
metallicity derived from the RGB bump are broadly consistent with what
we find from the location of the RC.  A further comparison of our
derived value of $M_{I,RC}$ with predicted values presented by
\citet{rej05} (their Fig. 22; based on models by \citet{piet04}) are
also consistent with a metal-poor population with an age of $\sim 8$
Gyr.

As an additional check, we compare our CMD with that for the Galactic
globular cluster NGC~362, studied as part of the $HST$ Treasury
project GO-10775 \citep{sar07, mf09}.  This cluster has a
metallicity of [Fe/H]$= -1.09$ (on the \citet{cg97} scale), and an
age of $\sim$10.5 Gyr \citep{mf09}, and therefore provides a good
reference for our field.  Figure 9 shows our M81 photometry compared
with the NGC 362 fiducial sequence derived from the HST Treasury
project data. The latter has been shifted by $\Delta(m_{F814W})$=13.12
and $\Delta(m_{F606W}-m_{F814W})$=0.05.  These are based on adopted
values of $(m-M)_o = 14.77$ and $E(B-V) = 0.037$ for NGC 362
\citep{sara09} and $(m-M)_{F814W} = 27.95$ and $E(F606W-F814W) =
0.080$ for M81.  We see good agreement between the location and shape
of the NGC 362 fiducial and the locus of M81 stars along the red clump
and the RGB. The color distribution of M81 RGB stars is slightly bluer
than the NGC 362 RGB fiducial, and is most evident within one
magnitude of the M81 RGB tip. This could be due to NGC 362 being
slightly more metal-rich ($\sim$0.05 dex) and somewhat older (1 to 2
Gyr) than the mean properties of the M81 field. Taken together, these
could account for a color difference of $\sim$0.03 mag in
$m_{F606W}-m_{F814W}$ on the RGB.

\section{Blue Stars}

Our CMD of the M81 halo field shows a (small) population of blue stars
with $m_{F606W}-m_{F814W}\approx0$, properties expected for young
($\approx10^8$~yr) stars, but not for the stars in the halo of a
spiral galaxy.  We have checked that these are genuine sources with
blue colors.  They are clearly point-like and not obvious background
galaxies, and are scattered throughout our ACS field.  We conclude
that these are likely individual, young stars associated with M81.
Furthermore, the rather sudden appearance of these objects at
$m_{F814W}\sim 26$ blueward of the dominant RGB suggests that these
are not RGB or RC stars in M81 scattered due to large photometric
errors.  An overlay of metal-rich (Z=0.008; [M/H]$\sim -0.4$)
isochrones from \citet{mar08} onto our CMD is shown in Fig. 10, and
suggests that these blue stars have ages between
$\approx200$-$400$~Myr.

What is the origin of these stars?  They have colors and magnitudes
similar to a stellar population seen in the CMD of an outer 
disk field in M81 \citep[Figure 3 in ][]{wil09}.  Potentially then, there is a
sparse outer disk in M81, which contributes a very small fraction of
the stars observed in our field.  If these stars are an extension of
the true outer disk of M81, then the disk plane extends out
$\approx$34-40 kpc from M81.

A second possibility is that these stars formed in the disk of M81 or
in one if its companions, and then were carried to this location along
with the HI tidal debris that is coincident with our field, likely as
a result of the most recent galactic interactions between M81, M82 and
NGC 3077, which occurred $\sim 220-280$ Myr ago \citep[][see also Weisz
\etal\  2008]{yun99}.     

The observed HI gas along the line-of-sight to our ACS field has a column density of 
only $N_{HI} \sim 10^{20}$cm$^{-2}$ \citep{yun94}.    Recent ($\sim 30-70$ Myr) star
formation has been found not only in the outskirts of M81 itself
\citep{bark09} but also in nearby tidal dwarf galaxies such as
Holmberg IX \citep{weisz08,sab08}, BK3N \citep{mak02}, and in other
nearby tidal debris \citep{dur04,dav08,chib09,mou10}.  However, in all of
these cases, the recent star formation has occurred in areas with
higher HI column densities ($N_{HI} > 4-8 \times 10^{20}$cm$^{-2}$)
than that observed in our field, and is thus consistent with the lack
of {\it recent} (i.e. $\tau \la 10^8$~yr) star formation in our field.
This is also consistent with the findings of \citet{may07}, who find
recent low-level star formation at levels $N_{HI} \sim
4\times 10^{20}$cm$^{-2}$).  We note that while recent episodes of star
formation in the outermost regions of late types galaxies have been
observed at column densities similar to that in our field \citep[$\sim
10^{20}$cm$^{-2}$; e.g.][]{dbw03,thil07}, we do not see evidence of
this in our (small) field.

We conclude that while we do see evidence for blue stars in our field,
these make only a small contribution to the stellar population in our
M81 field.

\section{Discussion and Conclusions}

The main result from our analysis of a field at a (projected) distance
of $18\arcmin$ (19 kpc) from the center of M81, just beyond the
suggested edge of the thick disk, is that the dominant population has
a mean age of 9$\pm$2 Gyr and a peak metallicity of [M/H]$= -1.15
\pm$ 0.11.   Below, we discuss these results in the context of the
formation history of M81, and compare with results from the spheroids
of other nearby galaxies.

The color distribution of stars on the RGB in our M81 field is
relatively blue and rather narrow.  We derived the metallicity
distribution of these RGB stars by comparing with theoretical
isochrones, and find a mean [M/H] of --1.15, with an intrinsic
dispersion of 0.37 dex as shown in Figure~8.  Our field is
dominated by stars that are more metal-poor than stars located closer
to the center of M81; indeed, our ACS field contained the most
metal-poor stars in any M81 field observed thus far.  In Figure~11, we
compare our MDF (for an assumed age of 9 Gyr) with that determined by
\citet{mou05} (for globular cluster-like ages; $\sim 12$~Gyr 
for a field that is also along the southwest minor axis, but somewhat
closer to the center of M81 with a projected distance of $\sim$14~kpc
(see Figure 1).  While the general shape for the MDF is similar in
both fields, the Mouhcine \etal\ field is clearly dominated by stars
which are more metal-rich by $\approx 0.4$~dex than the stars studied
here (where we have already accounted for the differences in the
assumed ages of the MDFs -- adopting a slightly older (by $\sim 3$
~Gyr) age would effectively {\it decrease} the mean metallicity of our MDF by $\sim 0.1$~dex; making 
the difference between the MDFs larger).  This metallicity difference is
not likely to be due to uncertainties in the colors or in the adopted
reddening values; to match a 0.4 dex difference in [M/H], our mean RGB
would need to be redder by $\sim 0.1$ in ($m_{F606W}-m_{F814W})$.
Similarly, the MDFs of \citet{mou05,mou06} are more metal-poor than
the outer (thick?) disk field presented by \citet{wil09}, a field
dominated by relatively old (2/3 of stars formed more than $\sim$ 8
Gyr ago) but relatively metal-rich stars.

This sequence of declining metallicity with location indicates that
either there is a strong metallicity gradient in the disk of M81, or
that our field (as well as the Mouhcine \etal \ field) samples a
different structural component of M81 than \citet{wil09}.  The
markedly lower metallicity between our field and that of the outer
disk of M81, combined with the relatively old ages inferred for these
stars (discussed below) and the lack of (much) recent star formation,
provide tentative evidence that our field is largely dominated by
stars in the {\em halo} of M81, rather than in a thick disk.

In a recent wide-field study of bright stars in M81's outer regions
\citet{bark09} suggest the presence of a faint, structurally distinct
component surrounding M81 with a mean metallicity [M/H]$\sim -1.1 \pm
0.3$; the large uncertainty is due to the rather limited part of the
RGB observable in their ground-based data.  This metallicity is
slightly higher than (but consistent with) the value found here \footnote{
it is important to note that \citet{bark09} assumed an age of
10 Gyr; if instead they had assumed 9 Gyr as we derive here, this
would increase their derived metallicity less than $0.1$ dex}.  Our field is
not located within the region surveyed by \citet{bark09}, but lies at
a deprojected distance of 34-40~kpc from M81, comparable to the distance of the
fields used for their metallicity estimate.  This extended component
has properties similar (but not identical) to either the Milky Way
thick disk or the halo, making the exact nature of this structure
somewhat uncertain.  Future deep observations of M81's outer regions
will be needed to help establish the true nature of this feature.

We can compare the MDF that we derive for our 20~kpc field in M81 with
the metallicities found for the spheroids/halos of other nearby
galaxies.  At galactocentric radii of $R<20-30$ kpc, similar to the
location of our M81 pointing, the spheroid of M31 is dominated by a
largely metal-rich ([Fe/H]$\sim -0.6$) population
\citep[\eg][]{dur01,kal06,gil07,rich08}.  The stars in the halo of the
Milky Way are metal-poor, with typical [Fe/H] values of
$-1.4$ to $-1.6$ for the inner halo (out to $\approx10-15$~kpc)
\citep{rn91,iv08} and $\approx-2.2$ beyond this \citep[][and
references within]{car07}.  The halo of the E/S0 galaxy NGC 3379 (at a
galactocentric distance of 33 kpc) shows a broad metallicity
distribution with a significant number of metal-poor ([M/H]$<-0.7$)
stars \citep{har07b}.  In contrast, the large elliptical galaxy NGC
5128 has a metal-rich halo with [M/H]$=-0.6$ to $-0.7$ at {\it
all} locations studied thus far, from 8-38 kpc
\citep{har99,hh00,hh02,rej05}.  Similarly, the elliptical galaxy NGC
3377 also shows a largely metal-rich population at small
galactocentric radii \citep{har07a}.

Taken together, large, nearby galaxies exhibit a wide diversity of
halo metallicities, showing no clear trends with parent galaxy mass or
morphological type \citep[although there are suggestions that halo
metallicity increases with galaxy mass, \eg ][]{mou06}.  Indeed, it
appears that variations in the metallicity of halo stars {\it even at
different locations within a given `halo'} could make any trends with
parent galaxy properties difficult to quantify.  For example, the
metallicity of stars in our single M81 field are similar to that of
the `halo' of NGC 3379, the `outer halo' of M31, and the `inner halo'
of the Milky Way (assuming $[\alpha/Fe]\sim +0.3-0.4$).  If we have
indeed sampled the halo of M81, an open question is whether an even
more metal-poor population exists at larger galactocentric radii; again, future studies of more distant M81 
field would be worthwhile.

While our CMD does not reach the main sequence turnoff region, the
most reliable age-sensitive feature in the CMD, it still holds clues
to the dominant age of the stars in this region of M81.  Formal fits
to the RC, RGB, and RGB bump give consistent estimates for the mean
age of the stars of $9\pm2$~Gyr.  The fact that we do not observe a
strong population of blue horizontal branch stars is also consistent
with these results.

We can compare the ages estimated for stars in the spheroids of
different galaxies, although it should be noted that these estimates
come from different techniques\footnote{while different studies assume
differing ages (\eg 12~Gyr or 14~Gyr) for the `old' populations, the
effect of these differences on metallicity are small, and do not affect
our general conclusions}.  The dominant population of our M81 field is
slightly younger but consistent with the mean ages (9.7, 11.0, and
10.5 Gyr, for fields at distances of 11, 21, and 35 kpc from M31) of
the M31 halo fields studied by \citet[][and references within]{bro08}.
The halo of M31 also harbors globular clusters as old as those in the
Milky Way \citep{brito09,ma09}; however, the presence or absence of
any radial age trends among this population is still largely an open
question.

In contrast, the stars in our M81 field appear to be slightly younger
(by $\approx2-3$~Gyr, but just within the age uncertainties), than the
oldest stars in the Milky Way halo \citep{mf09}, as our comparison
with the globular cluster NGC 362 illustrated above.   While there is
an age spread of a few Gyr among the Milky Way GCs with the inner halo
clusters being older in the mean than those of the outer halo
\citep{dot09}, the ages of globular clusters in M81 (estimated from
integrated spectra) are indistinguishable from Galactic globular
clusters \citep[within the sizeable uncertainties][]{sch02}.  It should
be noted that these clusters are all projected at smaller distances in
M81 than the location of our field.  We note that our data do {\it
not} rule out the presence of a minority population of older stars, as
ancient as the majority Galactic globular cluster system, in this
region of M81.

In the case of NGC 5128, \citet{rej05} investigated the ages of field
stars at a distance of 38~kpc using the magnitude of the RC, and found
that it is consistent with an old population with a mean age of $8\pm
3$ Gyr, very similar to the ages that we observe in the M81 halo.
 
Our age and metallicity results for halo stars in M81 are in broad
agreement with predictions from some $\Lambda$CDM models
\citep[\eg][]{bj05,font08}, where the outer ($R>10-20$ kpc) halos of
$L^*$ galaxies are expected to be formed from the accreted (over a
relatively long timescale) debris of somewhat lower-mass progenitor
galaxies than the more metal-rich systems that (for instance) resulted
in the formation of the more metal-rich Giant Southern Stream in M31
\citep{fard06,fard08,font08}.  In this scenario, the outer halos of
luminous galaxies like M81 will have more metal-poor and somewhat
younger combined stellar populations than that of the (slightly)
older, more metal-rich inner halo/spheroid.  Our results are
broadly consistent with these predictions.

\section{Summary}

We have presented deep photometry of a field located 20 kpc from the
center of M81, believed to be dominated by halo stars.  The resulting
CMD, based on HST/ACS imaging, reveals a relatively blue RGB and a HB
populated predominantly redward of the instability strip (i.e., a red
clump).  From the shape of the RGB, the magnitude of the red clump,
and the location of the RGB bump, we derive a mean metallicity of
[M/H] = --1.15 $\pm$ 0.11 and an age of 9 $\pm$ 2 Gyr for the dominant
population in this region of M81's halo.  This is the lowest
metallicity found for stars in any portion of M81, almost certainly
because our field is more distant than any other portion of M81
studied to date.  Our age and metallicity results are broadly consistent with
the predictions from $\Lambda$CDM models, which predict lower
metallicities and younger mean ages for outer halos of spiral galaxies
than for inner halos.  Future studies that probe the metallicities and
ages of stars even further out in the halo of M81 will allow for
better constraints on the early formation of this galaxy.

\acknowledgments

A.S. and P.R.D. were supported through funds provided by NASA through
grants (GO-10604.01-A and GO-10604.02.A, respectively) from the Space
Telescope Science Institute, which is operated by the Association of
Universities for Research in Astronomy, Incorporated, under NASA
contract NAS5-26555.  The authors would like to thank Andy Dolphin and
Ben Williams for their assistance with the photometric reductions with
DOLPHOT, and the anonymous referee for suggesting changes that improved the paper.    
We would also like to thank Tom Brown and Ben Williams for providing comments on earlier
versions of the manuscript, as well as Michael Siegel, Min Yun, Katie Chynoweth and Mustapha
Mouhcine for helpful discussions.

{\it Facilities:} \facility{HST}.

\clearpage

\includegraphics[width=6.5truein]{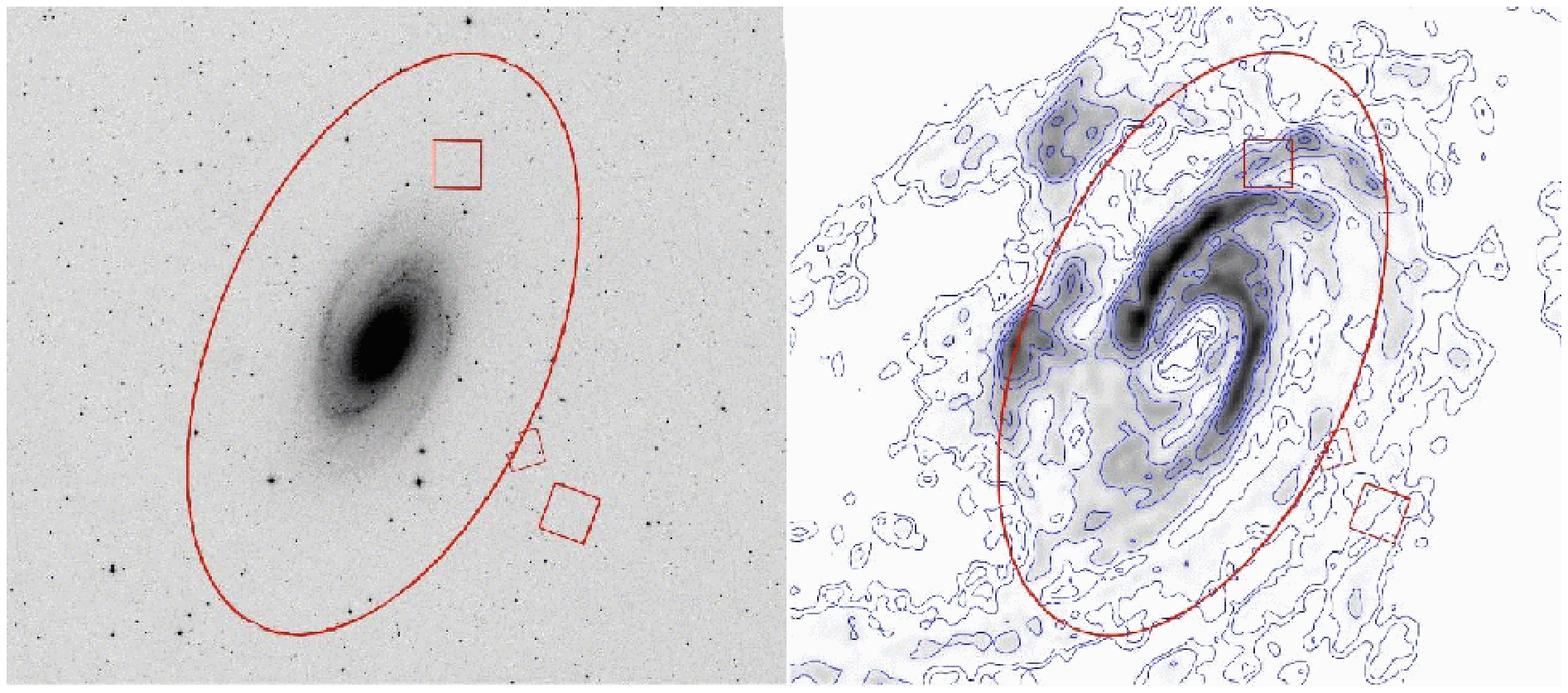}
\figcaption[fig1x.ps]{{\bf (Left:)} The location and field of
view of our ACS field (at lower right), superimposed on a red
Digitized Sky Survey image of M81.  The ACS field is located somewhat
beyond the WFPC2 field of \citet{mou05}, which is also shown.
The ellipse indicates the 22\arcmin ($\sim 25$ kpc) extent of M81's
thick disk as suggested by \citet{tik05}, inclined at 59\degr.
The box at upper center is the location of the deep ACS study by
\citet{wil09,angst} {\bf (Right:)} The same thick disk and ACS field
pointings as the figure on the left are shown, overplotted on the
greyscale HI map of the M81 system by \citet{yun94}, where the
contours represent $N(HI) = 0.5, 1.0, 2.5, 5.0, 7.5$ and $10 \times
10^{20}$cm$^{-2}$.  While our ACS field is superposed on some of the
HI in the system, the column densities at this location are lower than
that expected in order for star formation to occur.  Both images are
roughly 48\arcmin \ high and 57\arcmin \ wide; north is up, and east is to
the left.  \label{fig1}}

\includegraphics[width=5.5truein]{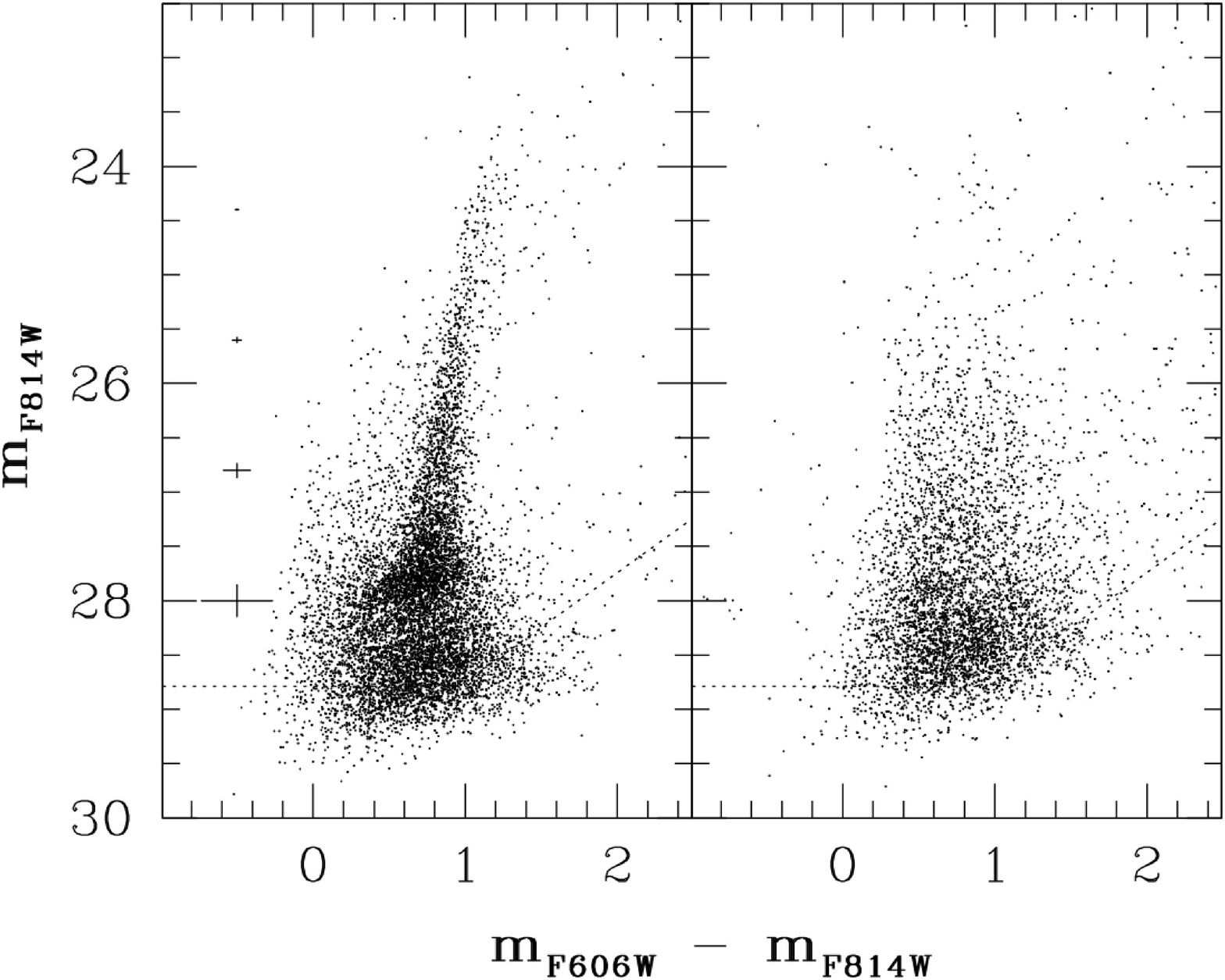}
\figcaption[fig2x.ps]{{\bf(Left:)} The $F814W$, $F606W - F814W$ 
color-magnitude diagram for all stellar sources measured in our ACS
field, after removal of objects which did not meet the selection
criteria ($\chi^2$, sharpness, and crowding) as defined in the text.
The error bars on the left denote the average errors in the photometry
based on the artificial star experiments.  The CMD is dominated by an
old red giant branch as well as a horizontal branch/red clump at
$m_{F814W}\sim 28$.  A sequence of blue ($m_{F606W}-m_{F814W} \sim 0$)
objects is also seen extending upward to $m_{F814W} \sim 26$.  {\bf
(Right:)} The CMD of all objects measured by DOLPHOT that were {\it
rejected} by our selection criteria.  The dashed lines in each figure
denote the $50\%$ completeness limits in both $m_{F606W}$ and
$m_{F814W}$.
\label{fig2}}

\includegraphics[width=5.0truein]{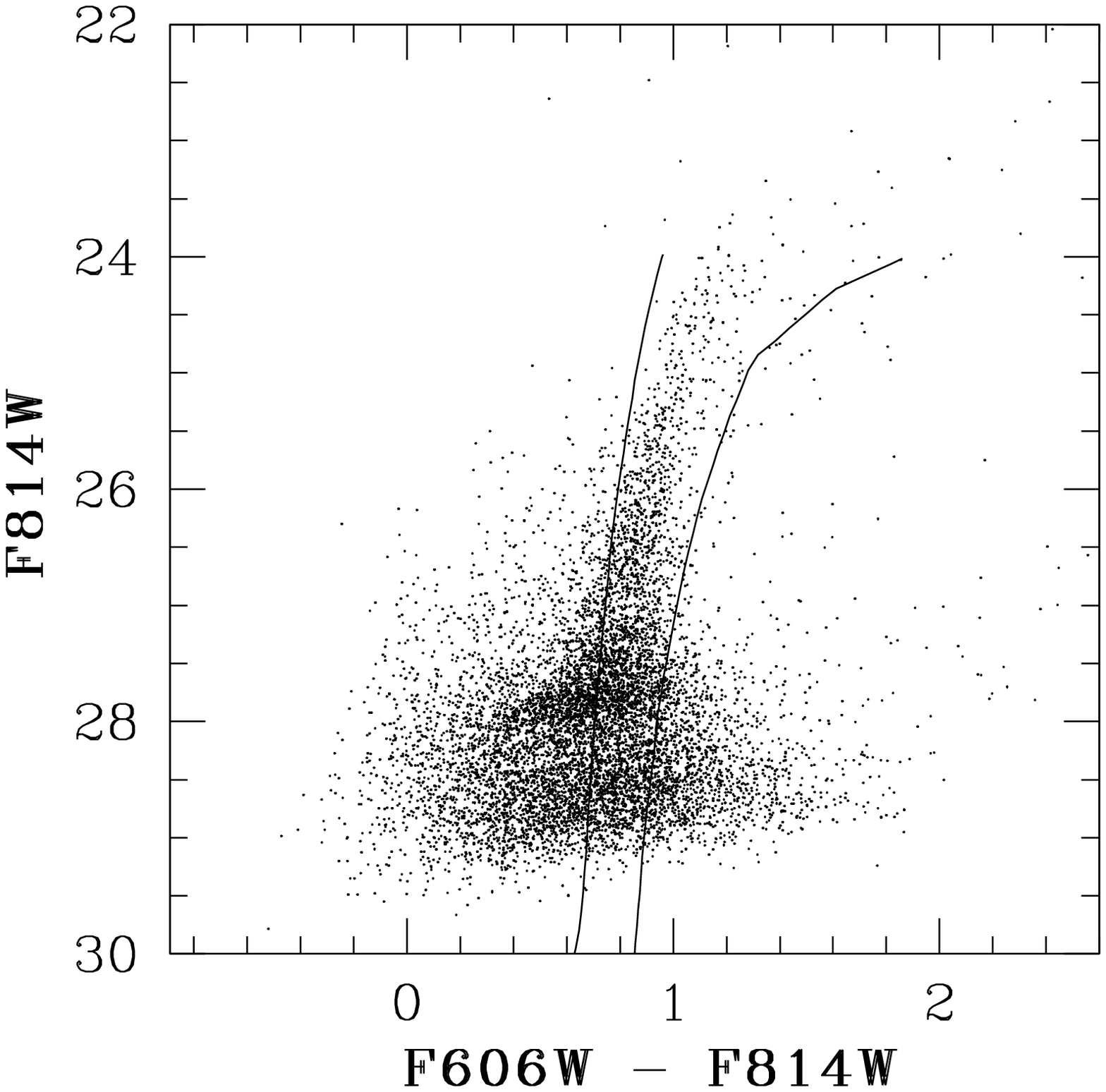}
\figcaption[fig3x.ps]{M81 halo CMD showing the loci used to define 
our RGB sample.  The 12.6 Gyr Girardi \etal~ (2002) Z=0.0001 (left) and
Z=0.004 (right) models have been shifted outward by 0.05 magnitude to account
for photometric spread.  The models were also shifted via a distance
modulus $(m-M)_{F814W}= 27.95$, and reddened by $E(F606W-F814W)=0.08$.
\label{fig3}}

\includegraphics[width=5.5truein]{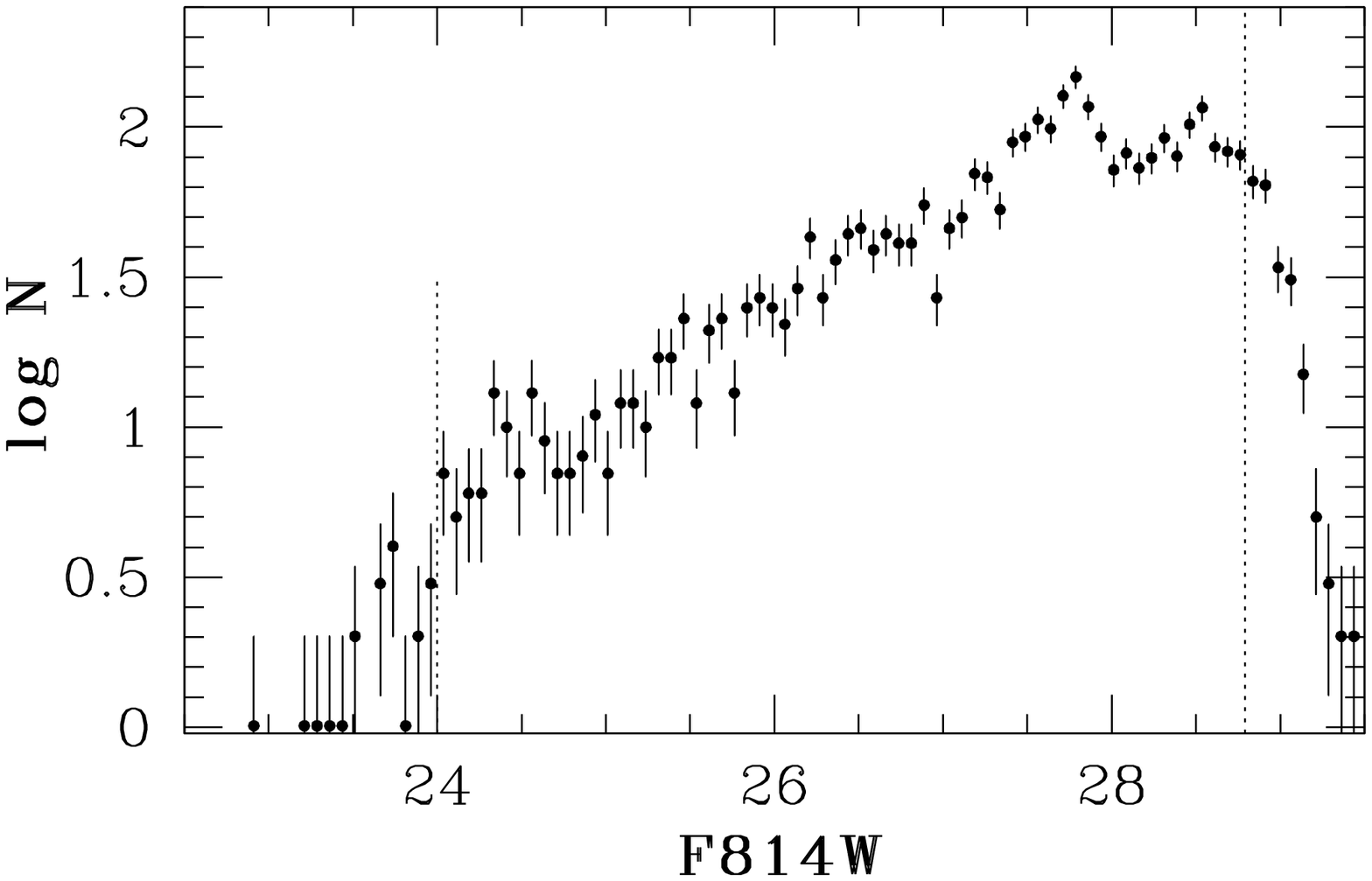}
\figcaption[fig4.eps]{$m_{F814W}$ luminosity function for all stars 
between the RGB loci in Figure 4.  No corrections for photometric
incompleteness have been made, nor have any (small) foreground star
contributions been removed.  The uncertainties are simply the
$\sqrt{N}$ statistics for the objects detected in each 0.075 mag bin.
The dashed line at $m_{F814W}=28.79$ denotes the 50\% completeness level of the photometry.
The solid line at $m_{F814W}=24.00$ represents the location of the RGB
tip.\label{fig4}}

\includegraphics[width=5.5truein]{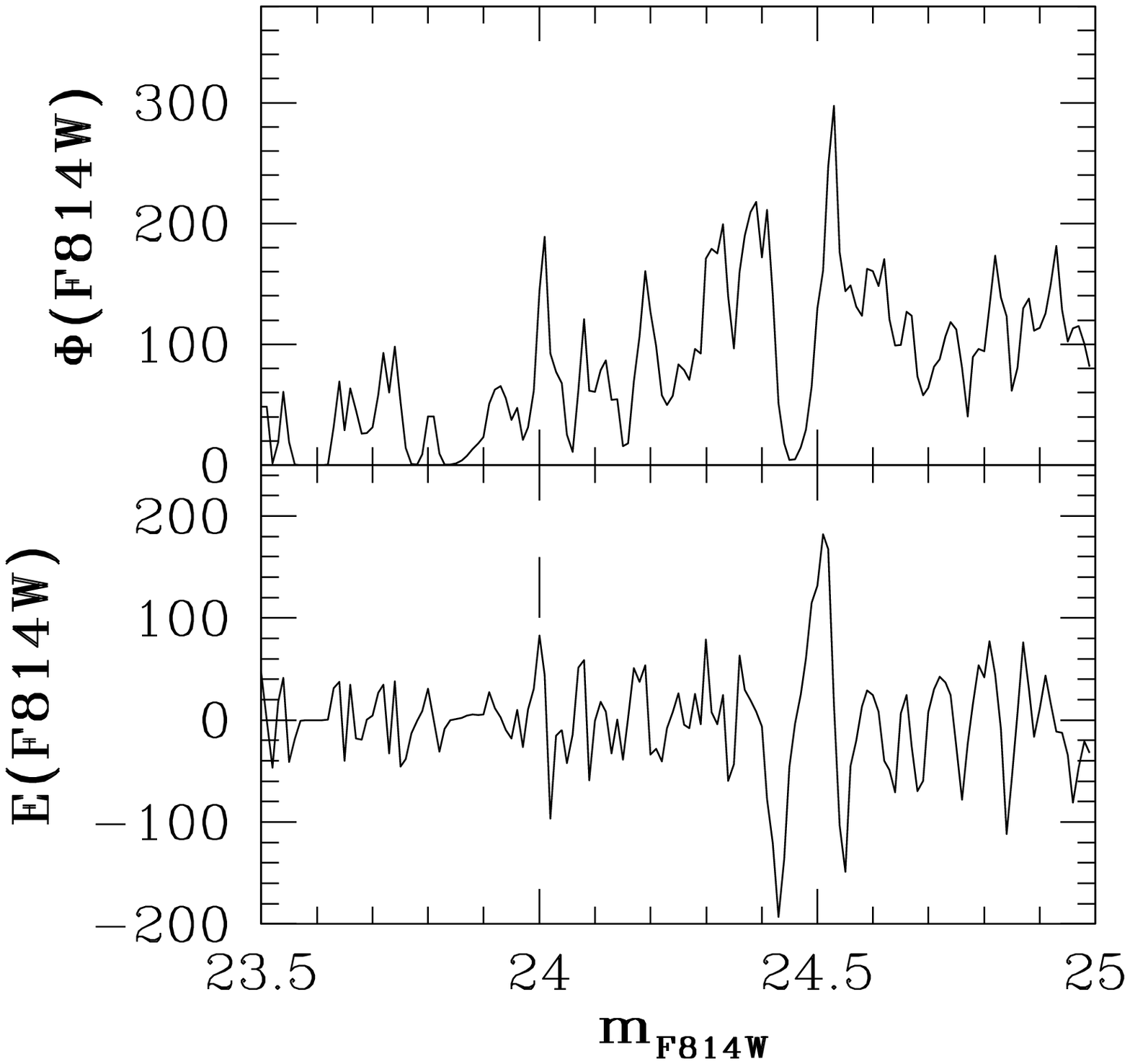}
\figcaption[fig5.eps]{The top panel shows a smoothed luminosity 
function formed by co-adding the Gaussian representation of stars in
our RGB subsample (see text).  The bottom panel is the result of
applying the Sobel edge-detection algorithm.  The peak at
$m_{F814W}$$_{TRGB} = 24.00 \pm 0.02$ (internal error only) is the
most likely location of the RGB tip.
\label{fig5}}

\includegraphics[width=5.5truein]{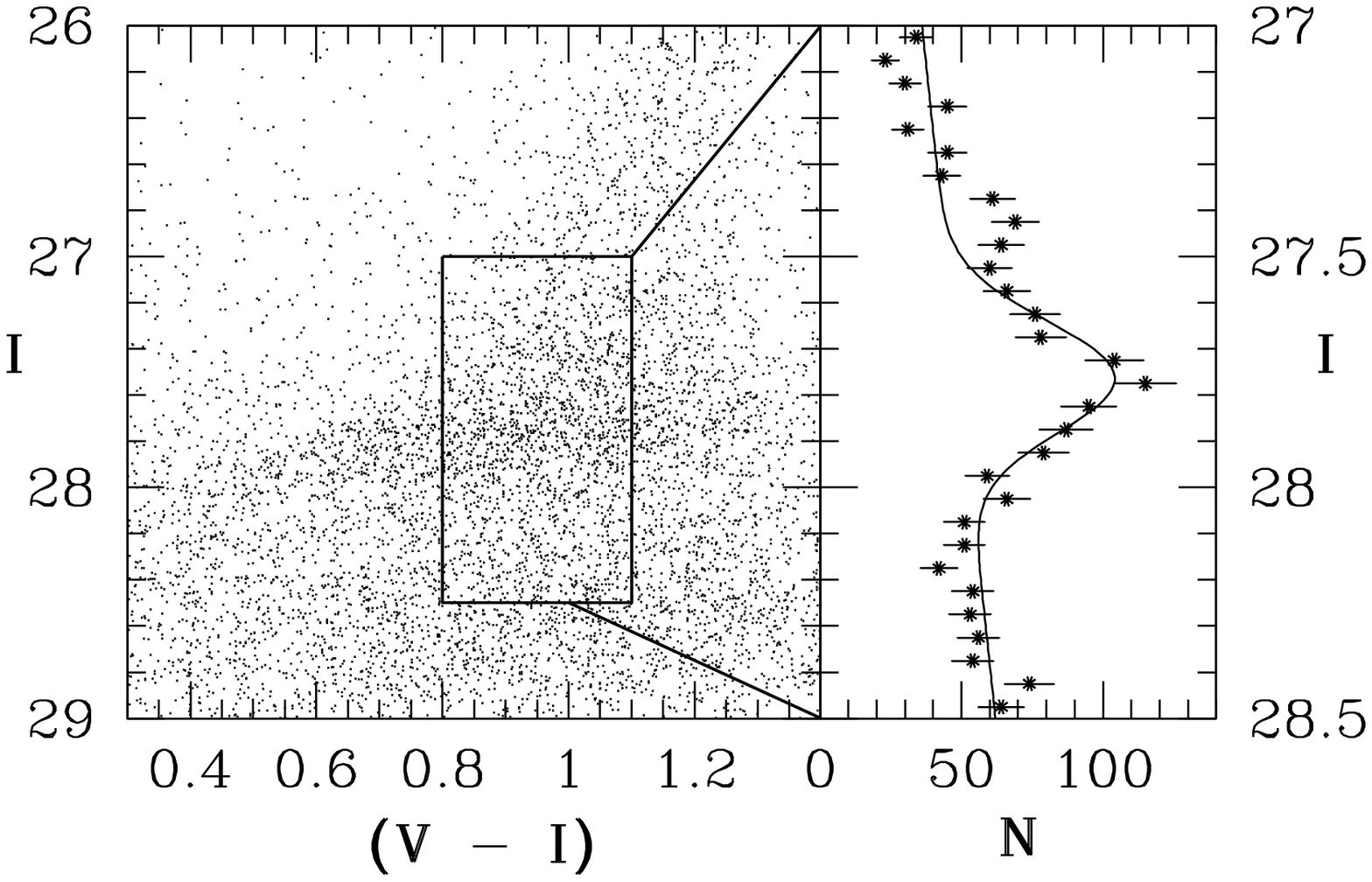}
\figcaption[fig6.eps]{{\bf (Left:)} A close up of the region around the red 
clump and horizontal branch in the $I$, $V-I$ color-magnitude diagram (transformed from the
ACS magnitudes using the \citet{sir05} transformations -- see text).
The box illustrates the region within which the red clump magnitudes
have been derived. {\bf (Right:)} The binned $I$ band luminosity function for
objects within the boxed region.  The LF has not been corrected for
photometric incompleteness.  The solid line shows the combined
linear background and best-fit Gaussian to the location of the red
clump, at $I=27.76\pm 0.04$, and a dispersion $\sigma=0.12_{+0.04}^{-0.03}$
\label{fig6}}

\includegraphics[width=5.5truein]{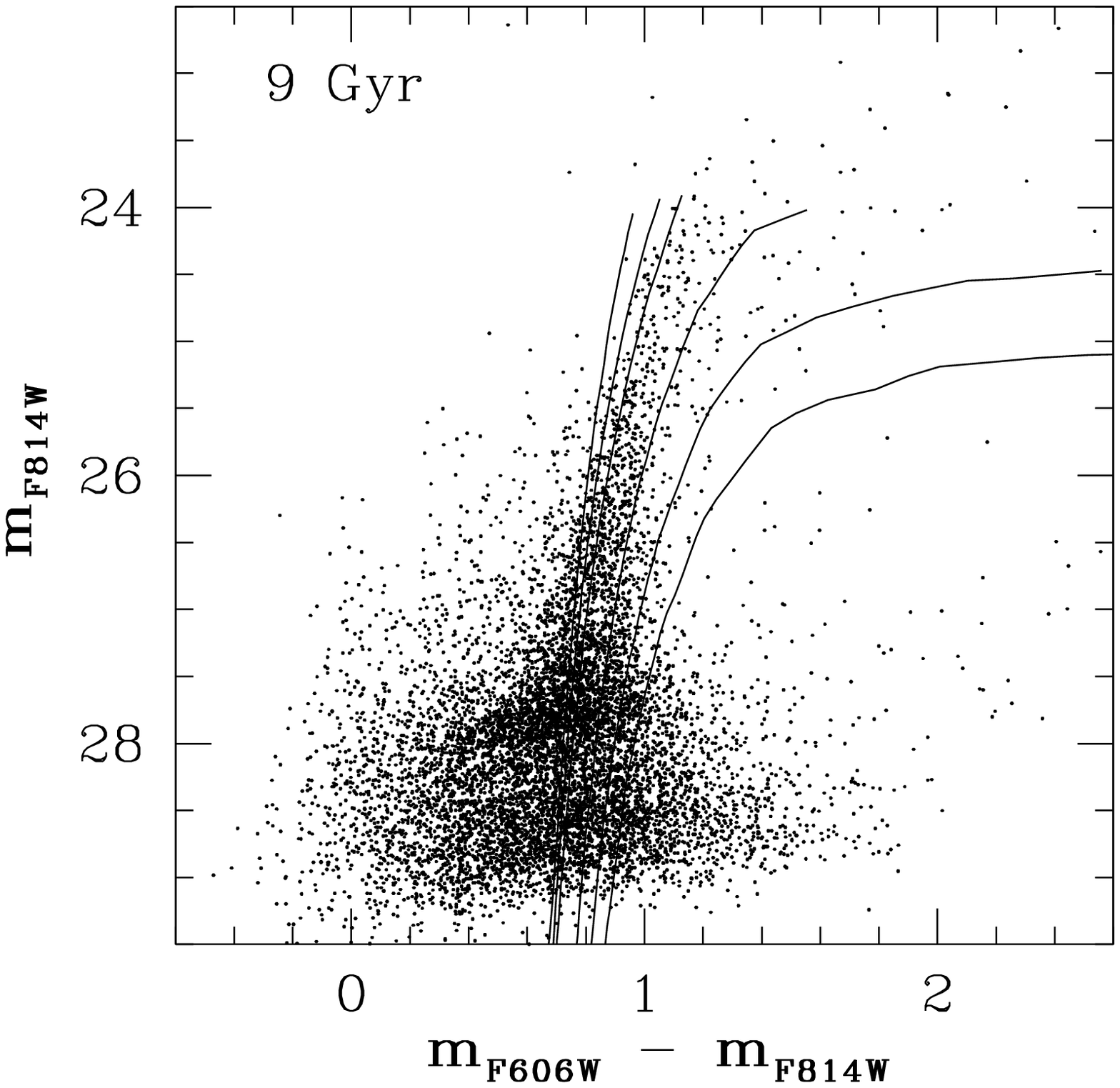}
\figcaption[fig7x.ps]{Our M81 field CMD overplotted with the 9 Gyr
RGB models from \citet{mar08}, for metallicities of (from left to
right) [M/H]$=-2.3, -1.7, -1.3, -0.7, -0.4, 0.0$.  All models have
been shifted by $E(F606W-F814W)=0.08$ and $(m-M)_{F814W}=27.95$.
\label{fig7}}

\includegraphics[width=5.5truein]{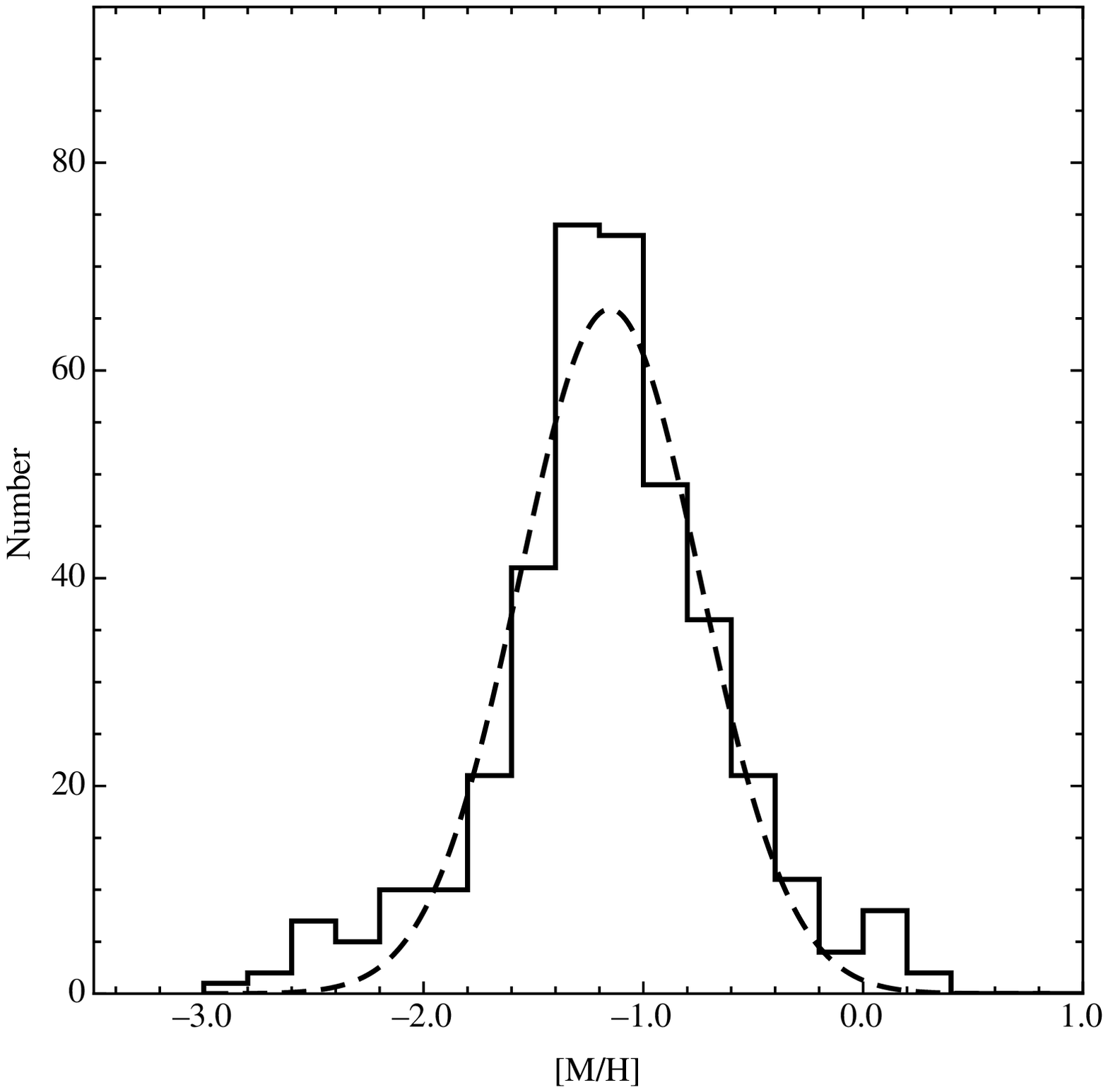}
\figcaption[fig8.eps]{A plot of the derived metallicity distribution 
function for stars in our M81 field.  The solid line is the binned histogram of 
metallicity determinations for RGB stars with $M_I$$<$--2.0 using 
the \citet{gir00} isochrones (see text). The dashed line is the Gaussian fit to the core of this
distribution.  We assume an age of log~$(\tau/\mbox{yr})=9.95$.  
\label{fig8}}

\includegraphics[width=5.5truein]{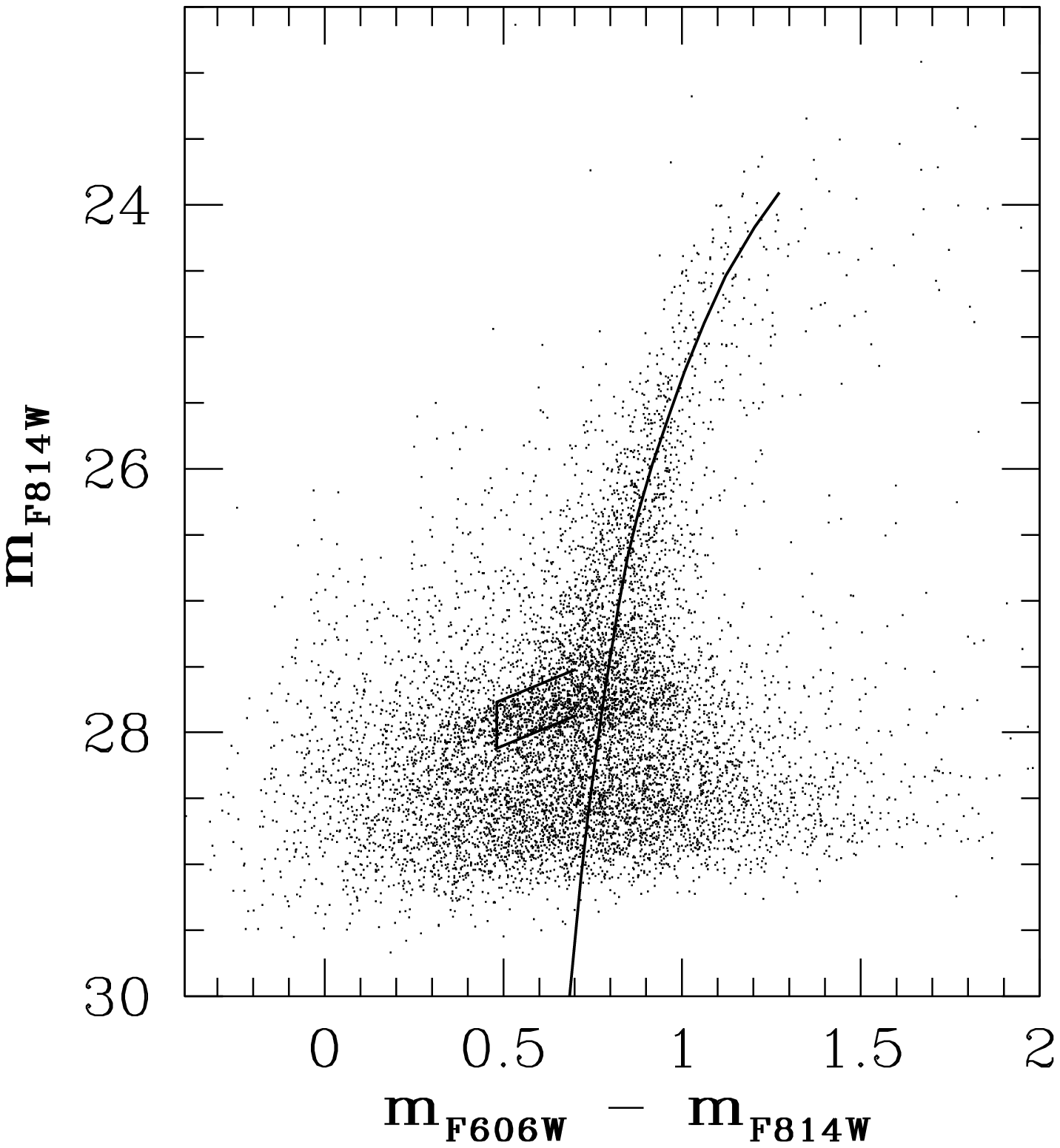}
\figcaption[fig9.eps]{The CMD of our M81 halo field, overplotted
with the fiducial of the Milky Way globular cluster NGC 362 from the
HST Treasury project, and the bounding box of the red HB.  The
fiducial has been shifted by $\Delta(m_{F606W}-m_{F814W})=0.05$ and
$\Delta m_{F814W} \sim \Delta I = 13.12$ to match the distance and
reddening of M81.  \label{fig9}}

\includegraphics[width=5.0truein]{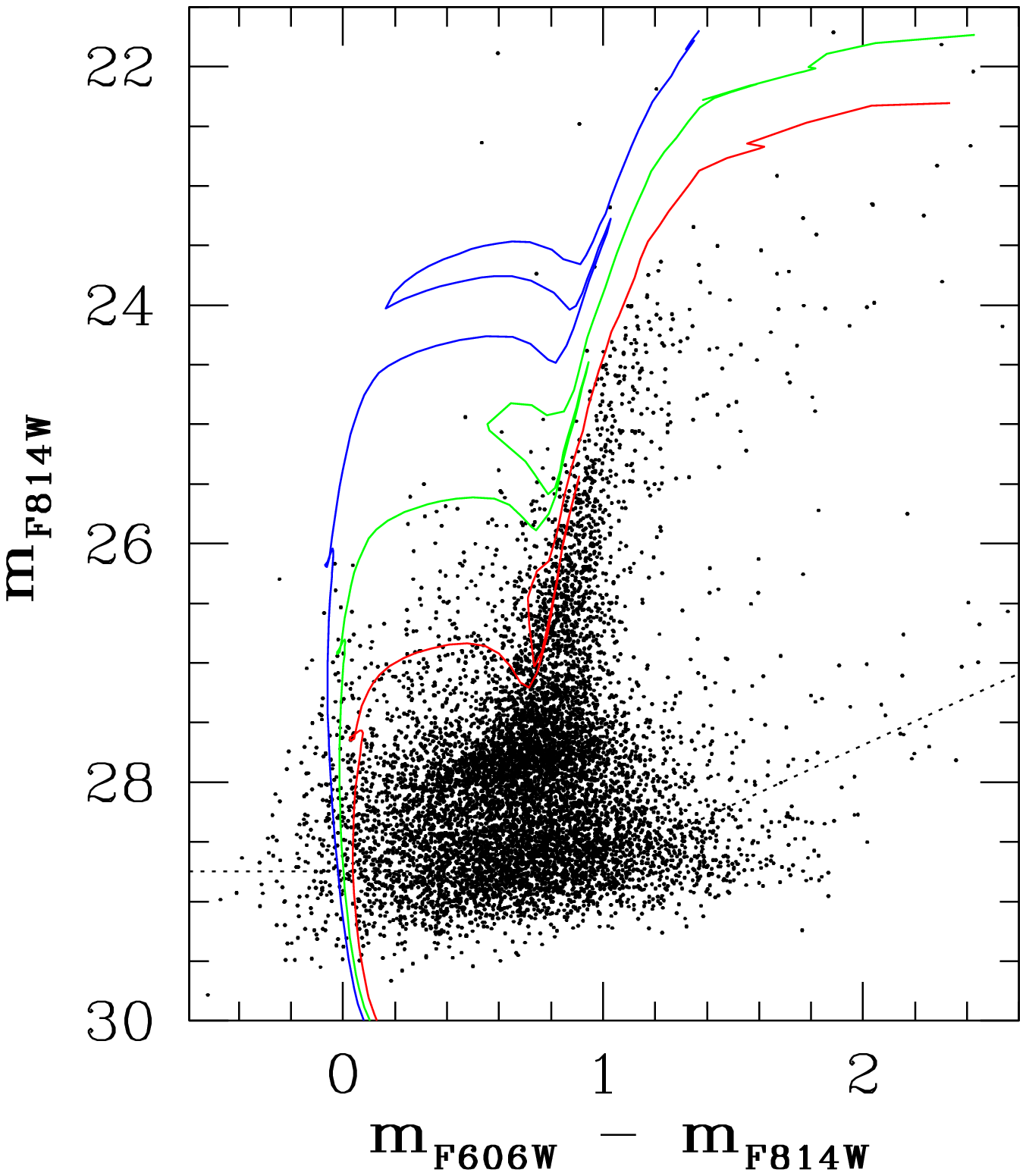}
\figcaption[fig10x.ps]{Our M81 field CMD overplotted with Z=0.008 isochrones from \citet{mar08}.  The ages are (from top to bottom) 100 Myr, 200 Myr, 400 Myr.   As before, all models have 
been shifted by $E(F606W-F814W)=0.08$ and $(m-M)_{F814W}=27.95$.
\label{fig10}} 

\includegraphics[width=5.5truein]{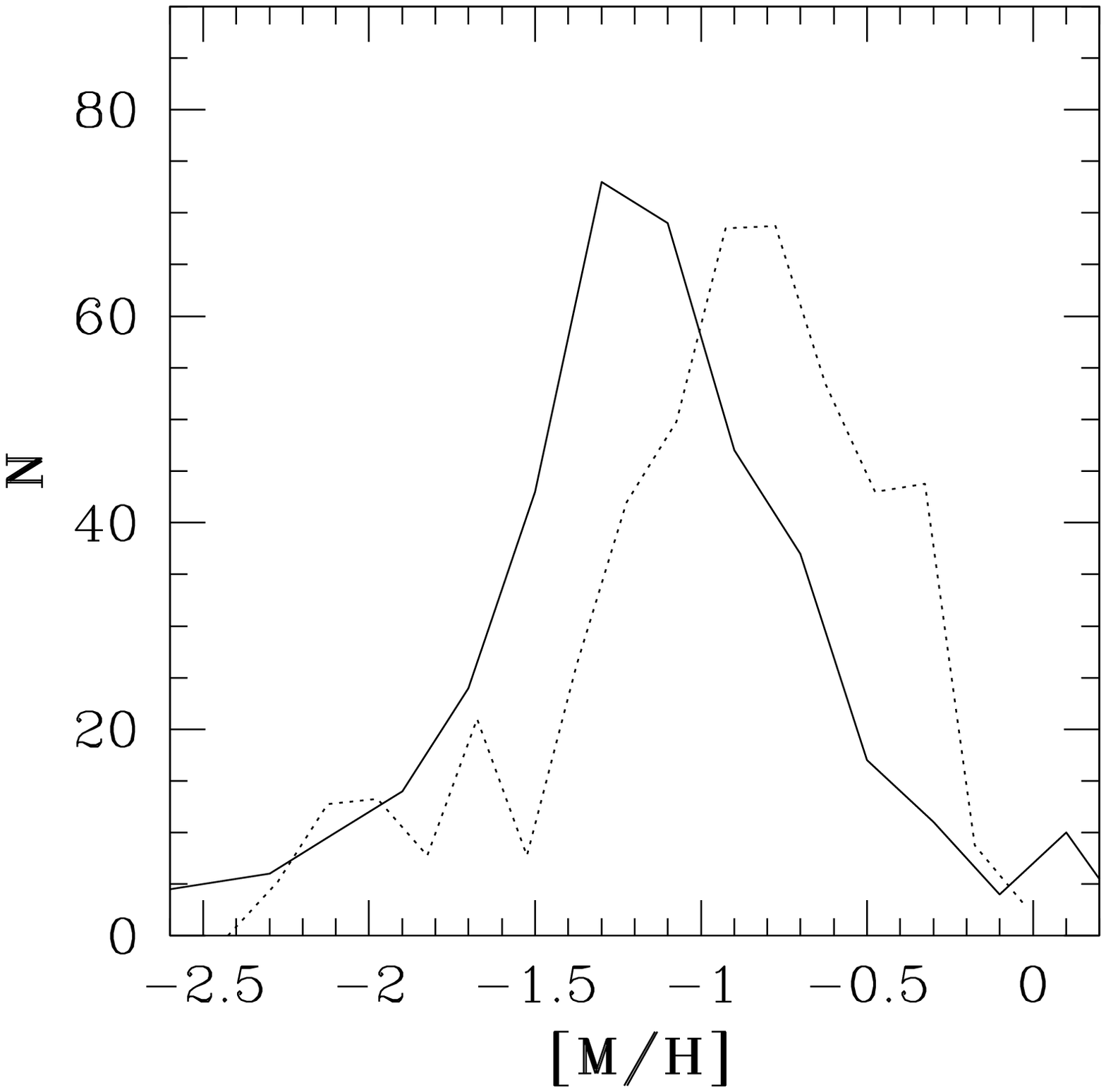}
\figcaption[fig11.eps]{Comparison between our MDF (for an age
of 9 Gyr) [solid line] and that of the MDF of an interior field from \citet{mou05} (for an assumed age of $\sim 12$ Gyr; dashed line)
\label{fig11}} \end{document}